# Deployment of Drone Base Stations for Cellular Communication Without Apriori User Distribution Information

Xiaohui Li[1]

1. School of Electrical Engineering and Telecommunications, University of New South Wales, Sydney 2052, Australia
E-mail: xiaohui.li@student.unsw.edu.au

**Abstract:** Drone base stations can provide cellular networks in areas that have lost coverage due to disasters. To serve the maximum number of users in the disaster area without apriori user distribution information, we proposed a 'sweep and search' algorithm to find the optimal deployment of drone base stations. The algorithm involves polygon area decomposition, coverage control and collision avoidance. To the best of our knowledge, this paper is the first in the literature that studied the deployment of drone base station without apriori user distribution information. Simulations are presented showing that the proposed algorithm outperforms the random search algorithm and the attractive search algorithm regarding the maximum number of severed users under the deployment of drone-BSs they found with a time limit.

## 1 Introduction

The terrestrial communication network may be partially or fully manufactured when some exceptional circumstances, such as earthquakes, floods, and debris flows occur. Meanwhile, ground transportations are often blocked in such emergency situations, which prevents emergency communication vehicles from getting close to the disaster area. In this case, high-altitude base stations can be used to provide emergency communications in the disaster area, where communications play a critical role in saving lives [1]. With recent advancements in drone technology, construct the high-altitude base stations by utilizing drones to carry the communication load for cellular networks has attracted considerable attention [2]. As a rapid solution to provide wireless connectivity, drone mounted base stations (drone-BSs) can assist cellular networks in cases of emergency communication, public safety communication and post-disaster rehabilitation [3]. For instance, with drone-BSs providing voice/data services, disaster victims can request help and report their locations, first responders can receive and relay critical information they need, hence boost the rescuing effort.

Unlike terrestrial base stations (BSs) and BSs mounted on ground vehicles, drone-BSs can be deployed in any location without being constrained by ground traffic conditions [4]. Obviously, drone-BSs should be deployed in locations where the maximum number of users can be covered. The backhaul transmission of drone-BSs can be achieved by wireless links to any of the functional core network resources such as satellites, nearby ground macro BSs, emergency communication vehicles or surviving base stations. The endurance (flight time) is one of the critical challenges to the feasibility of drone-BSs. In December 2017, a record of four hours and forty minutes in self-powered drone's endurance was claimed by a European drone manufacturer company 'Quaternium'[1]. Additionally, an architecture that consists of drones with switchable batteries and the ground drone bases for battery replacing can achieve persistent coverage to a specific location, by alternately deploying two or more drones.

In addition to degrading or destroying standard communication infrastructure, natural disasters often lead to geographical changes and chaotic movements of affected people. Therefore, drone-BSs are required to be rapidly deployed in the immediate aftermath of disasters to the operating environments that little may be known and may change unpredictably. In the context of a disaster scenario, an experiential victims' distribution of affected area may not reflect the real situation on the ground. Disasters such as earthquakes, floods and typhoons often affect wide geographical areas. Due to the nature of the disasters, the affected areas are usually convex. Meanwhile, most of the non-polygon areas can be transformed to polygon areas by polygonal approximation [5]. Thus, the operating areas considered in this paper are all convex polygons and contain no holes.

Among different generations of cellular networks including GSM/3G/LTE, one of the common functions of cellular BSs is to persistently broadcast system information on the forward channel [6-8]. Mobile stations (MSs) carried by network users consist of cell phones and other cellular module equipped devices. In the disaster area, MSs have not been covered by any operational base station will continuously perform an IMSI attaching procedure: scan the broadcast signal from cellular BSs, establish connections with the BS immediately after they received the broadcast signal [9]. Once the MS successfully connected to the cellular BS, the relative location of the MS will be stored in a database named 'visitor location register' (VLR), associated with the Mobile Switching Centre. Enabling cellular BSs to monitor the number of connected MSs at the real-time (e.g., by the length of the VLR). Since MSs are usually carried by network users, it is possible for drone-BSs to monitor the number of network users within its coverage

---

*This work was supported by the Australian Research Council.

1.https://newatlas.com/quaternium-record-endurance-drone-flight/52758/

during flight. Network users are usually concentrated at social attractions such as shopping malls, public transport junctions, residential and office buildings [10]. Besides, disaster victims tend to congregate rather than staying alone in many situations. Accordingly, user clusters will emerge, contain most network users in the affected area. The clusters' locations, however, are often unpredictable.

Thus, the objective that use drone-BSs to cover as many users as possible can be achieved by controlling a group of drone-BSs to find and cover the maximum number of user clusters in the operating region. To find locations of all the user clusters, firstly, the drone-BSs need to be deployed to detect every possible target area in the operating. In another word, to achieve sweep coverage in the operating region [11, 12]. Note that the 'coverage' here is different from the coverage of cellular BSs. Regarding coverage control, three types of coverage were defined in [11], i.e., barrier coverage [13], blanket coverage [14] and sweep coverage [15]. For barrier coverage and blanket coverage, sensing nodes are arranged statically to minimizes the probability of undetected ingress through the sensing barrier and maximizes the detection rate of targets appearing in a sensing area, respectively. Whereas sweep coverage aims to drive a group of sensing nodes so that every point in the sensing field is detected by some sensing nodes. In this paper, each drone-BS can be viewed as a sensing node while achieving sweep coverage. Specifically, drone-BSs detect the location of every user cluster by sensing the number of connected users while sweeping the entire operating region. During achieving sweep coverage in the operating area by a group of cooperative drone-BSs, collision avoidance becomes a problem needs to be considered. To solve the collision issue, the path planner of drone-BSs is required to guarantee collision avoidance while achieving the global objective [16]. In this paper, we only focus on the collision avoidance between drone-BSs. The collisions between drone-BSs and the other obstacles such as high buildings are neglected.

### 1.1 Related works

Recently, a considerable literature has grown up around the theme of drone-BSs in cellular networks. [17] presented a statistical generic air-to-ground RF propagation model for the drone-BS cell, facilitate the formulation of modeling problems of cellular networks with drone-BSs. The control problems of drone-BSs previously studied in, e.g., [2, 4, 18-22], are related to the 2D or 3D placement of drone-BSs. Where the objectives are to improve the coverage and power efficiency, maximizing the revenue and minimize the interference of drone-BSs.

[2] formulated the 3D placement problem of drone-BSs as a quadratically constrained mixed integer nonlinear optimization problem and proposed a numerical algorithm to maximize the number of covered users. [4] proposed a polynomial-time algorithm with successive placement, aims to minimize the number of required drone-BSs while ensuring the coverage of a group of mobile users. [18] proposed a method to find the 3D placement of drone-BSs in an area with different user densities using a heuristic algorithm, by decoupling the drone-BSs placement in the vertical dimension from the horizontal dimension. [19] proposed an optimal placement algorithm for drone-BSs that maximizes the number of covered users using the minimum transmit power. [20] optimized the locations of drone-BSs by brute force search to improve the throughput coverage and 5th percentile capacity of the network. The paper also discussed the relationship between the interference and separation of drone-BSs. [21] studied the dynamic repositioning trajectory optimization of non-holonomic drone-BSs based on the simple circular trajectory, aims to improve the power efficiency and spectral efficiency respectively. One of the limitations of existing literature, however, is much of the previous work assumes that the accurate locations of network users are apriori when studying the deployment problem of drone-BSs. Whereas the location information of network users is often unavailable, especially in the disaster area. [22] optimized the deployment of single and multiple drone-BSs to improve user experience in cellular networks based on greedy algorithms. The paper innovatively considered the inner drone distance constraint and drones' battery constraint during the optimization. The area of interest considered in [22] is the urban area that can be described by a street graph with apriori MS density function that is reflecting the traffic demand at certain positions on the street. MSs are assumed to be located near streets. Moreover, the 2D projections of drone-BSs are restricted to streets to avoid collision with buildings. In our paper, the operating area of drone-BSs is not restricted to the urban area and MSs are not limited to be located near streets. Besides, we assume that no any apriori MSs' location information is available, including the MS density function.

Papers relating to coverage control problem of mobile robots include [12, 23-27]. Where [12] proposed a decentralized control algorithm for mobile robots to accomplish sweep coverage based on simple consensus algorithm. [23] addresses the problems of barrier coverage and sweep coverage by self-deployed mobile robots based on the nearest neighbor rule and formation consensus. In contrast to the work [12, 23], we consider the control of drone-BSs is a centralized control problem as the drone-BSs have full ability to communicate with the control centers and all other drone-BSs, not only the neighbor drone-BS. We aim to develop a set of centralized control laws that steer a group of drone-BSs moving along given paths to achieve sweep coverage and, at the same time, search for the optimal placements that can serve the maximum number of users. Sweep coverage problems studied in [24-26] are related to complete coverage path planning problems. To solve this type of problems, a map of the operating region is required to be known or able to be constructed online for the path planner to generate robot paths that can completely cover the operating region. In this paper, we assume that the map of the operating region is apriori and partly adopt the coverage path planning procedure proposed in [27]. Accordingly, we proposed a centralized control strategy to achieve efficient sweep coverage by area decomposition and zig-zag sweeping path generates by the control center. Previous studies on collision avoidance between a group of mobile robots can be found in [16, 28-32]. Where [31] proposed a collision-free navigation algorithm for a non-holonomic robot in the unknown complex environment. [29] presented a real-time navigation law for mobile robots to avoid collisions based on sliding mode control. [16] introduced a reactive strategy for the navigation of a mobile robot in prior unknown environments with moving and deforming obstacles. [32]

described a method of collision avoidance for UAV based on simple geometric approach. In this paper, we partly adopted the method proposed in [32] to achieve collision avoidance between two drone-BSs.

### 1.2 Contributions

In this paper, we study a challenging problem: deploying a limited number of drone-BSs in a relatively wide disaster area, with no or very few operational terrestrial base stations, aims to serve as many users as possible, assuming no apriori user distribution information. We address the problem by a method we called 'sweep and search'. The method involves deploying drone-BSs to achieve sweep coverage by area decomposition and zigzag path planning. Firstly, convert the disaster area into a convex polygon as the operating area for drone-BSs, then perform polygon area decomposition to ensure that each sub-area is assigned to a drone-BS. Secondly, deploy every drone-BS to its assigned sub-area, sweep the area on the optimal direction follows the zigzag pattern, search around to find the location that can cover the entire user cluster once it detected a concentration of users, then continue to finish the sweeping. The proposed algorithm offers a promising solution to find multiple user clusters and related optimal placement of the drone-BSs. To the best of our knowledge, the concept of sweeping and searching user clusters to server the maximum number of users by drone-BSs has not been discussed in the literature.

### 1.3 Organization of the paper

The rest of this paper is organized as follows. Section 2 presents the control model of the drone-BSs' deployment problem. In section 3, we introduce our 'sweep and search' method. Some simulation results and discussion are presented in section 4. Finally, section 5 concludes the paper.

## 2  System Model

We consider a team of drone-BSs engaged in searching for cellular users in a geographical area, where most of existing ground cellular BSs were manufactured because of a disaster. With the objective to serve as many users as possible, provide GSM/3G/LTE mobile network to responders and victims in the disaster area.

### 2.1  Air-to-ground channel model

The drone-BSs' deployment problem can be simplified by decoupling the deployment in the vertical dimension from the horizontal dimension. Firstly, we optimize the altitude of drone-BSs by analyzing the air-to-ground path loss of drone-to-user links. According to the air-to-ground channel model proposed in [17], the probability of having a line-of-sight (LoS) connection between a drone-BS and an MS plays a vital role in modeling the air-to-ground path loss. The probability is formulated by

$$P_{LoS} = \frac{1}{1 + a\exp(-b(\frac{180}{\pi}\phi - a))} \quad (1)$$

Where $a$ and $b$ are constant values that depend on the environment (urban, rural, and so forth) and $\phi$ denotes the elevation angle of the drone-BS, equals to $\arctan(h/r)$, where $h$ and $r$ are the drone-BS' altitude and the horizontal distance between the drone-BS and the MS, respectively. The probability for the drone-to-user links to have a non-line-of-sight (NLoS) is $P_{NLoS} = 1 - P_{LoS}$. According to equation (1), the probability of having LoS link increases as $\phi$ increases. It implies that $P_{LoS}$ will increase by increasing the drone-BS' altitude, with the horizontal distance $r$ fixed.

Path loss of radio signals consists of two groups: long-term variations like free space propagation loss and small-scale variations such as losses due to scattering and shadowing. However, the planning of BS deployment is mainly dealing with the long-term variations [33]. Hence, in this paper, we focus on the mean path loss rather than the random behaviors of the radio channel. Then the mean path loss model can be formulated by [17]

$$L_P(h,r) = 20\log(\frac{4\pi f_c \sqrt{h^2 + r^2}}{c}) + P_{LoS}\eta_{LoS} + P_{NLoS}\eta_{NLoS} \quad (2)$$

Where $L_P$ is the path loss (dB) for the MS, neglecting the MS's altitude and antenna height. $c$ is the speed of light (m/s), $f_c$ is the carrier frequency (Hz). Moreover, $\eta_{LoS}$ and $\eta_{NLoS}$ are the mean additional losses in dB corresponding to LoS and NLoS connection, respectively, depending on the environment [17]. The coverage of the drone-BS is disk-like with the drone-BS' projection on the ground as the center and a coverage radius of $R_d$. To guarantee a certain quality of service (QoS) for the MSs located at the boundary of the coverage disk, the path loss at the coverage boundary must be less or equals to a specific threshold $L_{th}$, i.e. $L_P <= L_{th}$. Accordingly, the drone-BS' coverage radius has

$$R_d = r\,|_{L_P(h,r) = L_{th}} \quad (3)$$

For two environments (urban and suburban) and a given air-to-ground path loss threshold $L_{th} = 100 dB$, the coverage radius versus the altitude of a drone-BS is shown in Fig. 1.

As can be seen from Fig.1, for a specific environment (urban, suburban) and a given path loss threshold, there is an optimal altitude for drone-BSs regarding coverage maximization. This is because of the following nontrivial trade-off: decreasing free space path loss requires lower altitude of the drone-BS, which in turn reducing the possibility of having LoS links for MSs.

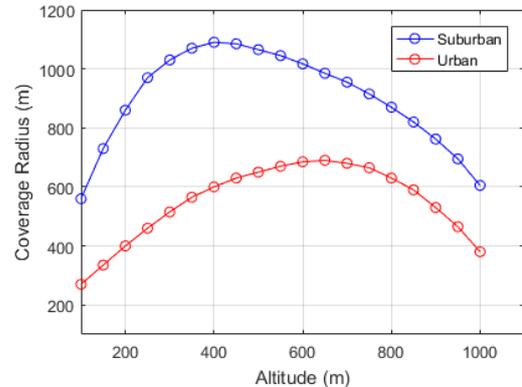

Fig. 1: Coverage radius VS drone-BS' altitude

The optimal altitude $h^*$ can be obtained by solving

$$h^* = h\,|_{\partial r/\partial h = 0} \quad (4)$$

Then the corresponding coverage radius $R_d$ can be solved by substituting $h = h^*$ into equation (1) and (2). In the following parts of this paper, we only focus on finding the optimal horizontal deployment of drone-BSs.

## 2.2 Deployment problem

We control a fleet of drone-BSs $D = \{D_1, D_2, \cdots, D_n\}$ to find their optimal placement. Where $n$ donates the number of drone-BSs. The operating area is modeled as a convex polygon in the Cartesian coordinate plane, with an area of $A_P$. The area of each drone-BS' coverage region is $A_d = \pi R_d^2$. Throughout this paper, we assume

$$A_p \gg n A_d \quad (5)$$

It implies that the drone-BSs need to efficiently explore a relatively large operating area, to achieve the best coverage. During the flight, each drone-BS can monitor the number of users within its coverage region after every sampling time $T_s$. Each drone-BS is configured to share its location and velocity to all other drone-BSs through the control center. With the presence of measuring error and transmitting delay, the shared location of a drone-BS should be treated as a disk with a radius of $r_s$, rather than an accurate coordinate.

Assume that users within the operating area are either stationary or of low-mobility. A specific number of user clusters are randomly distributed in the operating area with fixed center locations. Note that users are not necessarily to be within the user clusters. The number of users stayed out of any clusters $u_O$ and the number of the clustered user $u_C$, however, satisfy the following assumption

$$u_O \ll u_C \quad (6)$$

Which means user clusters contain most of the users within the operating area. Additionally, for any user cluster, we assume

$$d_{i,j} \leq R_d, \forall i, j \in U_C, i \neq j \quad (7)$$

Where $U_C$ is the set of all users within the cluster, $i$ and $j$ are two different users belong to $U_C$, and $d_{i,j}$ donates the distance between them. Equation (7) shows that a drone-BS can entirely cover any single user cluster. Therefore, the target of covering as many users as possible can be realized by utilizing drones-BSs to find locations of user clusters and then deploying the drones-BSs to the optimal locations that can entirely cover each cluster.

## 3 Proposed Algorithm

In this section, we describe our 'sweep and search' algorithm. The objective for drone-BSs is to accomplish sweep coverage and to locate as many user clusters as possible. Meanwhile, find the optimal placement for drone-BSs to cover every user cluster completely. Since user clusters are usually relatively sparse in the operating area, it is important for drone-BSs to thoroughly explore the environment to detect potential targets (i.e., user clusters). Therefore, we propose a 'sweep and search' solution consisting of two stages:

**a.** The control center first determines the operating area for drone-BSs as a convex polygon, then computes the optimal sweep direction and the decomposition of the convex polygon and assigns each generated sub-area to one drone-BS.

**b.** Deploy each drone-BS to sweep the assigned sub-area thoroughly, follows the zigzag pattern. Especially, search around the place where the concentration of users is detected to find the placement that can cover the entire user cluster. Then continue to finish sweeping the entire operating area.

### 3.1 Area decomposition and path planning

To thoroughly explore the operating area and avoid collisions between drone-BSs, we first decompose the operating area into several sub-areas. To decompose the area systematically, we adopt the polygon decomposition algorithm proposed in [27] because of its generality and simplicity. The inputs of the algorithm are the convex polygon $P$ with an area of $A_P$, the number of sub-areas $s$ that depends on the number of available drone-BSs, and the proportions indicate the area of $P$ that should be assigned to each sub-area. The outputs of the algorithm are the optimal sweep direction and a set of sub-areas $\{S_1, S_2 \cdots S_s\}$ with their areas equal to $\{A_1, A_2 \cdots A_s\}$, which satisfy

$$A_i = p_i A_P, i = 1, 2 \cdots s \quad (8)$$

$$\sum_{i=1}^{s} A_i = A_p \quad (9)$$

The proportions $\{p_1, p_2 \cdots p_s\}$ can be determined by relative capabilities of drone-BSs, such as remaining battery life.

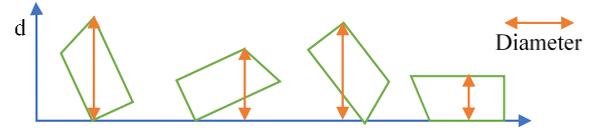

Fig. 2: Rolling the polygon to find the diameter function

Firstly, we create the diameter function $d(\theta)$ by rotating the polygon with the angle $\theta \in [0, 2\pi)$. Where diameter $d$ refers to the height difference measured between the highest and lowest points on the boundary of the rotating polygon [27], as shown in Fig. 2. Assume the angle that gives a minimum diameter $d_{\min}$ is $\theta_{opt}$, i.e.

$$d_{\min} = d(\theta)|_{\theta=\theta_{opt}} \quad (10)$$

Next, we compute the optimal sweep direction, which minimizes the number of turns needed along the zigzag pattern. Since turning a drone-BS is more time-consuming than letting it follows a straight line, by minimizing the number of turns we can reduce the total mission time. According to [27], for a vehicle that sweeping the entire polygon follows a zigzag pattern, the optimal sweeping direction is vertical to the direction gives the minimum diameter of the polygon. Finally, we divide the original polygon along the optimal sweep direction so that the area of each part is as same as calculated by equation (8). This can be done by sweeping a line along the polygon with the direction gives the minimum diameter, slicing the polygon once the area of the part on one side of the line equals to any $A_i$ that has not been cut yet [27]. After all, the minimum diameter directions of all sub-area will be the same as the minimum diameter direction of the original polygon. Hence, the obtained sweep direction is also the optimal sweep direction of all sub-areas (see Fig. 3). Fig. 4 shows the corresponding zigzag sweeping path in sub-area $A_3$ as an example. The summarized area decomposition algorithm can be seen in Table 1.

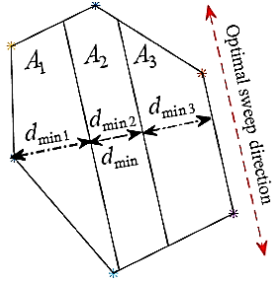

Fig. 3: Area decomposition

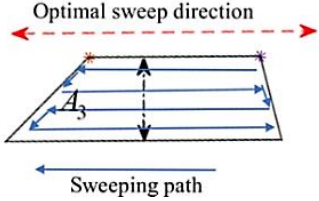

Fig. 4: Zigzag sweeping path in sub-area $A_3$

Table 1: Area decomposition algorithm

| Area Decomposition Algorithm |
| --- |
| 1: Create the diameter function of the polygon. |
| 2: Find the optimal sweep direction from $d_{\min}$. |
| 3: Slice the polygon with the divide lines parallel with the optimal sweep direction to match the pre-defined sub-area proportions. |

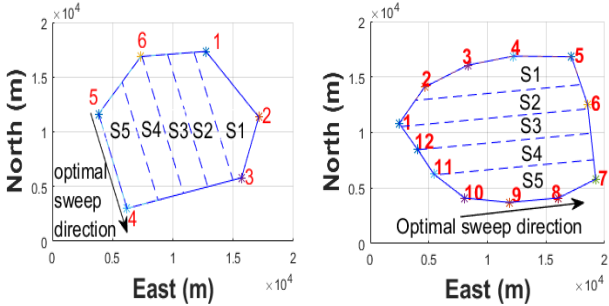

Fig. 5: Polygon area decomposition examples

Fig. 5 shows two examples of the polygon area decomposition results, with the number of the original polygon's vertex equals to 5 and 12, respectively. Both original polygons were divided into 5 sub-areas $\{S_1, S_2, S_3, S_4, S_5\}$ with the same proportions of 0.2. The black arrows show the optimal sweep directions.

### 3.2 User cluster searching

To efficiently accomplish the sweep coverage of the operating area, each drone-BS is allocated to a specific target area generated by the area decomposition process. We now describe how a single polygon area can be surveilled by one drone-BS in a way that guarantees 100% found of the best deployments corresponding to all the user clusters in the area. As discussed previously, when the drone-BS flies into the operating area with a fixed altitude, its coverage region on the ground moves with it. The coverage region is disk-like with a fixed radius of $R_d$, with the drone-BS' ground projection as the center. Once a drone-BS arrived its assigned operating area, it will continuously sweep the area until accomplished 100% sweeping coverage. The sweep follows zigzag pattern with the optimal sweeping direction obtained by the area decomposition algorithm in section 3.1.

During the flight, users located within the coverage will automatically connect to the drone-BS. By preventing the speed of drone-BS from too high, the time for building the connection can be neglected, because supporting users with high mobility (e.g., vehicle-based users) has been considered throughout the design of modern cellular networks [34]. The drone-BS senses and records the number of connected users $n_d$ in every sampling time $T_S$. Once it detected a significant increase of $n_d$ above a threshold $\varepsilon_{th}$ at the time $t_a$, i.e.

$$\Delta n_d > \varepsilon_{th}, \ t = t_a \qquad (11)$$

it will then recognize that the coverage region has been overlapped with part of a user cluster. Here, we define $t_a$ as the 'encounter moment'. The threshold $\varepsilon_{th}$ must be carefully set to find a balance between the false alarm ratio and the miss ratio. Starting from the point that an overlap has been detected, we want to find out how to change the drone-BS' location in some neighborhoods around to completely cover the user cluster while cover as many users around the cluster as possible.

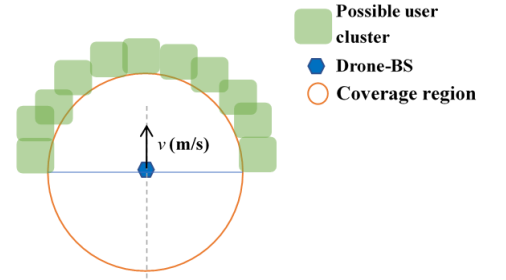

Fig. 6: Possible user cluster locations of the 'encounter moment'

Fig. 6 shows a snapshot of the 'encounter moment'. As illustrated in Fig. 6, the user cluster could be located around any part of the 'front half boundary' (corresponding to the drone-BS' moving direction) of the drone-BS' coverage region. Note that the user cluster can be of any shape, rather than the one shows in Fig. 6. Since the largest diameter of any single user cluster are smaller than the coverage radius of the drone-BS (see equation (7) and Fig. 2), the best placement of drone-BS to cover the entire user cluster can be obtained by steering the drone-BS to move along the 'front half boundary' of the coverage region at the 'encounter moment', as shown in Fig. 7. Note that the search path in Fig. 7 is the referential path that ignored the drone-BS' minimum turning radius. We can get the actual path by smoothing the above search path. Additionally, multiple solutions exist on the search path for covering the entire user cluster. By recording and comparing $n_d$ of each solution, the optimal deployment location relating to the user cluster for the drone-BS to cover the maximum number of users can be found. The optimal deployment will be immediately shared with all other drone-BS, and the user cluster will be labeled as 'detected'. The main idea is to avoid repeatedly searching an area by more than one drone-BS, especially for those user clusters located on both sub-areas. The coverage regions of surviving ground base station will also be labeled as 'detected area' where the inside user clusters will be ignored. The sweep procedure will be interrupted at each 'encounter moment' by above search procedure. At each breakpoint, the drone-BS' location $(x_0, y_0)$ and velocity $v_0$ will be remembered. As shown in Fig. 7, the drone-BS returns to $(x_0, y_0)$ when finishing the search procedure, it will then continue the sweep procedure with the velocity equals to $v_0$. Moreover, once a drone-BS stopped working during the sweep and search procedures, the drone-BS that next to it should take over the operating area of the failing drone-BS. By simply moving the dividing line to cover the remaining undetected area assigned to the failing drone-BS [27].

When drone-BSs finished searching, all user clusters located in the operating region will be found. Each user cluster will have one corresponding optimal location for deploying the drone-BS. The control center will rank the optimal locations based on the number of served users, then deploy drones-BSs to these optimal locations. If the number of user clusters is larger than the number of available drone-BSs $n$, then only the top $n$ optimal locations will be deployed by drone-BSs. Overall, the task of serving the maximum number of users will be accomplished by deploying all available drones-BSs to the optimal locations.

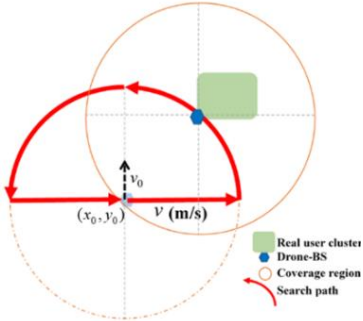

Fig. 7: Search path for finding the optimal deployment

### 3.3 Collision avoidance

As discussed in the previous section, we focus on avoiding collisions between drone-BSs. Drone-BSs are configured to share their locations and velocities to each other, the shared locations of drone-BSs are disk-like with a radius of $r_s$. We now partly adopt the method proposed in [32] to solve the collision avoidance problem. When two drone-BSs are getting closer, the minimum distance passed by each other should be larger than a specified minimum separation distance $d_{safe}$ to avoid collisions between them. There should be

$$d_{safe} > 2r_s \quad (12)$$

The pass distance can be found by the Point of Closest Approach [32]. As shown in Fig. 8, the pass distance vector $\vec{d}_p$ is defined as:

$$\vec{d}_p = \hat{c} \times (\vec{d} \times \hat{c}) \quad (13)$$

Where $\vec{d}$ is the relative distance vector and $\hat{c}$ is the unit vector in the direction of the relative velocity vector $\vec{c}$ from drone-BS 'B' to drone-BS 'A'. The magnitude of $\vec{d}_p$ is the pass distance $\|\vec{d}_p\|$. Naturally, we have the pass distance vector $\vec{d}_p$ and the relative velocity vector $\vec{c}$ are orthogonal:

$$\vec{d}_p \cdot \vec{c} = 0 \quad (14)$$

The time for drone-BSs to the closet approach $\tau$ satisfies:

$$\vec{d}_p = \vec{d} + \vec{c} \cdot \tau \quad (15)$$

Combine (13) and (14) we can get:

$$\tau = \frac{\vec{d} \cdot \vec{c}}{\vec{c} \cdot \vec{c}} \quad (16)$$

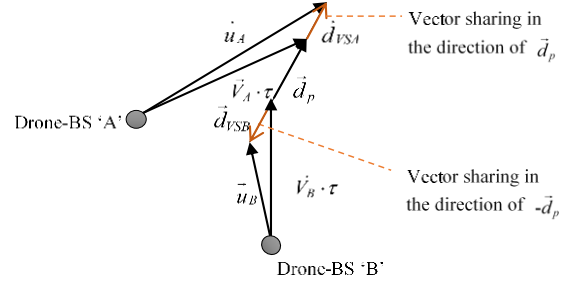

Fig. 8: Relative movement of two drone-BSs

By defining $d_{margin}$ as the margin pass distance, we have

$$d_{margin} = \|\vec{d}_p\| - d_{safe} > 0 \quad (17)$$

is the condition of no collision. If $d_{margin} < 0$, the control input $\vec{u}_A$ and $\vec{u}_B$ should be taken to enhance $d_{margin}$ by driving two drone-BSs to the opposite directions in parallel with $\vec{d}_p$, as shown in Fig. 9.

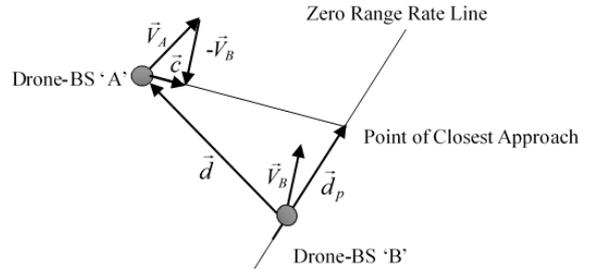

Fig. 9: Control input illustration

Where $\vec{d}_{VSA}$ and $\vec{d}_{VSB}$ are the vectors on the directions of $\vec{d}_p$ and $-\vec{d}_p$, respectively. The enhanced margin pass distance $d^*_{margin}$ becomes:

$$d^*_{margin} = \|\vec{d}_p\| + \|\vec{d}_{VSA}\| + \|\vec{d}_{VSB}\| - d_{safe} \quad (18)$$

By predefining $d^*_{margin}$ to a specific value that is larger than zero. $\vec{d}_{VSA}$ and $\vec{d}_{VSB}$ can be calculated as:

$$\vec{d}_{VSA} = \frac{d^*_{margin} \cdot |\vec{V}_B|}{|\vec{V}_A| + |\vec{V}_B|} (\frac{\vec{d}_p}{|\vec{d}_p|}) \quad (19)$$

$$\vec{d}_{VSB} = \frac{d^*_{margin} \cdot |\vec{V}_A|}{|\vec{V}_A| + |\vec{V}_B|} (\frac{\vec{d}_p}{|\vec{d}_p|}) \quad (20)$$

Finally, we can get the control input $\vec{u}_A$ and $\vec{u}_B$:

$$\vec{u}_A = \vec{V}_A \cdot \tau + \vec{d}_{VSA} \quad (21)$$

$$\vec{u}_B = \vec{V}_B \cdot \tau + \vec{d}_{VSB} \quad (22)$$

$$\vec{u}_A, \vec{u}_B \in \vec{U} \quad (23)$$

Where $\vec{U}$ sets the control limit. The collision between drone-BSs can be prevented by applying control input $\vec{u}_A$ and $\vec{u}_B$ at the time point when the condition of no collision (17) was found to be not holding.

## 4 Simulation Results

To assess the performance of the proposed algorithm, we simulated a team of 5 drone-BSs searching a $10km \times 10km$ quadrangle area with no operational ground base station. To generate the user distribution in the area, we first generate the locations of user clusters' centers through two dimensional uniform random processes. After that, we generate the locations of the clustered users and the locations of non-clustered users by two independent Poisson point processes (PPPs), with the user densities $\lambda_c$ and $\lambda_{nc}$, respectively. For simplicity, all user clusters are set to be disk-like, with the same radius $R_c$. The simulation parameters are as shown in Table 2. Drone-BSs have no prior information on the distribution of user clusters and the other users.

Table 2: Simulation parameters

| Parameter | Value |
|---|---|
| Drone-BS flight speed $v$ | 10 m/s |
| Sampling time $T_s$ | 0.5 sec |
| User cluster radius $R_c$ | 250 m |
| Coverage radius $R_d$ | 500 m |
| Clustered user density $\lambda_c$ | 0.02 users/$m^2$ |
| Non-clustered user density $\lambda_{nc}$ | 3e-6 users/$m^2$ |
| Threshold of user increase $\varepsilon_{th}$ | 10 users |

The performance of the proposed algorithm was compared to that of the random search algorithm and the attractive search algorithm. In random search algorithm, the drones-BSs move with random initial directions, and only change directions when their coverage regions hit the area boundaries or the other drones-BSs' coverage regions. In attractive search algorithm, the movement of drone-BSs is guided by inertia and their own best-known position where the maximum number of users can be served. Compare with the random search algorithm, for the drone-BSs in the attractive search algorithm, the attraction of their own best-known position can prevent them from missing the user clusters them encountered. All simulations were run for 50 times.

Fig. 10 and Fig. 11 shows how the number of severed users under the deployment of 5 drone-BSs found by different algorithms increases with time, for the searching area with 2 and 5 user clusters, respectively. As can be seen from both Fig. 10 and Fig. 11, the number of severed users under the deployment found by the proposed sweep and search algorithm exceeds that of the other two algorithms when the search time is above a particular value. The result could be attributed to that the proposed algorithm can always achieve a 100% sweep coverage of the operating area within a specific time, while the other two algorithms may leave parts of the area undetected, thus may miss some user clusters. By comparing Fig. 10 and Fig. 11 we can find the advantage of proposed algorithm becomes more evident for the operating area with fewer user clusters, which is more difficult for the random search to find user clusters.

## 5 Conclusion

In this paper, the deployment of drone-BSs over a disaster affected area without apriori user distribution information is investigated. By utilizing drone-BSs to detect the users' distribution during flight, we proposed a 'sweep and search' approach that offers a promising solution to find all user clusters in the operating area and related optimal deployment of drone-BSs that can cover as many users as possible. The approach consists of area decomposition, sweep coverage and collision avoidance. Simulations have shown that the proposed algorithm outperforms the random search algorithm and the attractive search algorithm regarding the maximum number of severed users under the deployment of drone-BSs they found with a time limit. The user distribution information detected by drone-BSs can also be used for rescue by disaster response organizations. In the future study, the collision avoidance between drone-BSs and other obstacles can be taken into consideration. Constraints such as the flying height of the drone-BSs can be considered. Future work can also extend to the cases with non-stationary users.

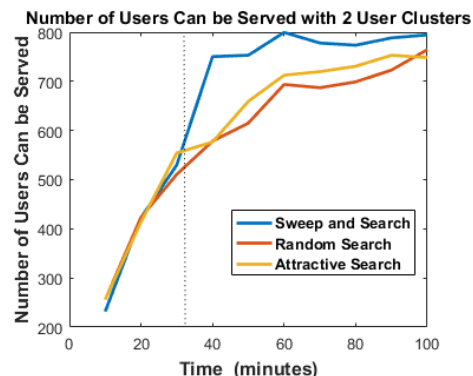

Fig. 10: Number of severed users VS search time for different algorithms, with 5 drone-BSs and 2 user clusters

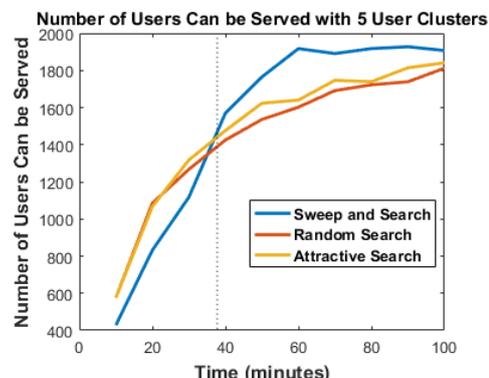

Fig. 11: Number of severed users VS search time for different algorithms, with 5 drone-BSs and 5 user clusters